\documentclass[11pt]{article} 
\usepackage[T2A]{fontenc}
\usepackage[utf8]{inputenc}
\usepackage[english]{babel}
\usepackage[round]{natbib} 
\usepackage{aas_macros}

\usepackage{adjustbox}
\usepackage{titling}
\usepackage{fancyhdr}

\usepackage[a4paper,margin=1in]{geometry} 
\usepackage{authblk} 
\usepackage[colorlinks=true, linkcolor=black, citecolor=black, urlcolor=blue, filecolor=blue,breaklinks=true]{hyperref} 
\usepackage{setspace} 
\usepackage{lineno} 
\usepackage{csquotes} 
\usepackage{ragged2e} 
\usepackage[capitalise, noabbrev, nameinlink]{cleveref}
\crefdefaultlabelformat{#2\textbf{#1}#3} 
\Crefname{figure}{\textbf{Figure}}{\textbf{Figures}} 
\Crefname{section}{\textbf{Section}}{\textbf{Sections}} 
\Crefname{table}{\textbf{Table}}{\textbf{Tables}} 

\usepackage{graphicx}
\graphicspath{{./main_figs/}{./supplement/suppl_figs/}} 
\DeclareGraphicsExtensions{.pdf,.jpeg,.JPG,.png,.PNG, .eps, .tiff}
\usepackage{subcaption} 
\DeclareCaptionLabelFormat{bold}{\textbf{(#2)}} 
\usepackage{pgffor} 
\usepackage{alphalph} 
\usepackage[labelfont=bf, textfont=bf, singlelinecheck=off, textfont=footnotesize]{caption} 
\usepackage{booktabs}
\usepackage{multirow}
\usepackage{rotating}

\makeatletter
\renewcommand{\AB@affillist}{}
\renewcommand{\AB@authlist}{}
\makeatother

\title{The orbits of visual binary and multiple stars obtained by 
the Apparent Motion Parameters method during the last 40 years}

\author[1,*]{L.G.~Romanenko}
\author[1]{O.V.~Kiyaeva}
\author[1]{I.S.~Izmailov}
\author[1]{N.A.~Shakht}
\author[1]{D.L.~Gorshanov}
\affil[1]{The Central Astronomical Observatory of the Russian Academy of Sciences at Pulkovo}

\affil[*]{\textit{e-mail:}\underline{lrom1962@list.ru}}
\date{January 2023} 

\begin{document}

\setstretch{1.15} 

\maketitle

\thispagestyle{fancy}

\begin{abstract}
Summed many years of work at Pulkovo, the orbits of 67 wide pairs 
of visual double and multiple stars (included in 64 systems) which 
were obtained by the Apparent Motion Parameters (AMP) method are 
presented. This short arc determination orbit method is based on 
the most reliable astrometric and astrophysical data corresponding 
to one instant of time. The rest of the observations accumulated in 
the world serve to control the quality of the orbit and refine some 
parameters. All early determined AMP-orbits were compared with new 
observations, some of them were recalculated, new ones were added. 
For the stars of Pulkovo program of observations with a 26-inch 
refractor, the Gaia DR2 data were analised. Based on these data, the 
orbits of 16 stars were calculated. In 20 cases from 67, the 
quasi-instant motion according to the Gaia DR2 data at the instant 
2015.5 contradicts the motion according to all-world observations. 
A possible reason is the presence of inner subsystems. The 
orientation of the obtained orbits in the galactic coordinate 
system is also given.
\end{abstract}

\section*{Introduction}

Double stars were first discovered by William Herschel at the end of 
the 18th century. In the 19th and 20th centuries, they were actively 
discovered by John Herschel, Wilhelm and Otto Struve, Paul Couteau 
and many other researchers. For stars with orbital periods of no more 
than 100-200 years, reliable orbits were determined, which made it 
possible to determine the masses of stars and derive fundamental laws 
relating the masses, spectra, luminosities and parallaxes of stars. 

Wide, slowly circulating binary and multiple stars have been deprived 
of the attention of researchers, since from the moment of their discovery 
to the present time, observations cover a small arc of the orbit. However, 
such stars may turn out to be outer pairs of open or not yet discovered 
multiple systems. Namely, towards multiple systems, the components of 
which are being discovered more and more, the interest of researchers 
has now shifted - see works ~\citep{1994AJ....107..306H,  
2015A&A...574A...6A, 2021Univ....7..352T}. Their study ultimately carries 
information about star formation and the dynamic evolution of subsystems 
in our Galaxy. Therefore, the problem of determining the orbits of poorly 
studied wide binary stars remains relevant. 

Observations of double stars have been a traditional theme of the Pulkovo 
Observatory since its opening in 1839. Therefore, when a 26-inch refractor 
appeared in Pulkovo after the Great Patriotic War, the priority task was 
to resume observations of wide binary stars in accordance with the 
capabilities of this telescope ($\rho>3''$).

Especially for determining the orbits of such stars, the Apparent Motion 
Parameters method (AMP) was completed, see ~\citep{1980AZh....57.1227K}. 
This method was previously used to determine the orbit of an artificial 
satellite of the Earth from one photograph with many exposures 
~\citep{1973AZh....50.1298K}. 

\section{Application of the AMP-method}

The AMP- method is designed to determine the initial orbits of wide visual 
binaries with a long period of revolution in position and velocity at one 
point in time based on the results of observations obtained by various 
available methods. These are the parameters of the apparent relative motion 
(AMPs) at one instant $T_\circ$: the distance between the components ($\rho$), 
the position angle ($\theta$), the visible relative motion ($\mu$) and the 
position angle direction of the visible motion ($\psi$), as well as the radius 
of curvature ($\rho_c$). The most accurate AMPs are obtained from homogeneous 
observations (basis) made with a single telescope to eliminate instrumental 
systematic errors (see, e.g., series of photographic observations with a 
26-inch refractor of the GAO RAS for 1957–2007: ~\citep{Pu2014,Pu2016}. 

In addition, the following data are necessary. They are: the parallax $p_t$ 
(to relate linear and angular quantities),  the relative radial velocity 
of the components $\Delta{V_r}=V_{rB}-V_{rA}$ (km/s), obtained from 
spectroscopic observations (to calculate the space velocity vector of 
the satellite relative to the main star), and an estimate of the sum 
of the masses of the components $\Sigma{M}$, according to the data on 
physical properties of these stars.

If it is possible to determine all five parameters, including the radius 
of curvature, then the distance between the components $r$ in astronomical 
units can be calculated by the formula:

\begin{equation}
\label{one}
 r^3=k^2\frac{\rho\rho_c}{\mu^2}|\sin
(\theta -\psi)|
\end{equation}

where $k^2=4\pi^2\Sigma{M}$ is  dynamic constant if distance is measured 
in au, time in years, mass in $M_\odot$. 

Then we get two position vectors that correspond to the position of the 
secondary component symmetrically with respect to the picture plane, and, 
consequently, two orbits ($\pm\beta$). Here $\beta$ is the angle between 
the spatial position of the satellite and its projection on to the picture 
plane, which can be calculated by the formula: 

\begin{equation}
\label{two} \beta=\pm\arccos{(\rho/(p_tr))}
\end{equation}

Sometimes, in agreement with the entire series of observations (see {\it 
Washington Double Star Catalog} --- WDS, ~\cite{WDS}) one can choose one 
solution.

If the radius of curvature cannot be determined, then the distance between 
the components, we obtain according to the condition necessary for an 
elliptical orbit: 

\begin{equation}
\label{three} \frac{\rho}{p_t}\leq{r}<\frac{2k^2}{v^2},
\end{equation}

where $v$ is the modulus of spatial velocity in au/year. In this case, we 
get a family of orbits. Each orbit of the family is characterized by the 
angle $\beta$.

If it is impossible to estimate the radius of curvature from a short arc 
of homogeneous observations, but the entire series of available observations 
reflects a nonlinear elliptical motion, then a more correct orbit can also 
be chosen from the family in agreement with the entire series of observations. 

In order to obtain high-accuracy AMPs, at the Pulkovo Observatory much 
attention was paid to obtaining a homogeneous long-term series, from which 
the relative motion is determined more confidently. However, for some stars 
it was necessary to break the homogeneity and use additional observations 
from the WDS catalog.

The AMP orbits for the two stars included in this work (HIP 12706 and HIP 
33287) required minor corrections and were refined using the differential 
correction method using Tokovinin's ORBITX program ~\citep{1992ASPC...32..573T}. 
For these stars, Table 2 compares the AMPs obtained using the basis indicated 
in the table and the AMP ephemeris corresponding to the orbits refined using 
the ORBITX program. 

Sufficiently accurate radial velocities are not available in the literature 
for all the stars in our program. Sometimes this parameter also had to be 
selected in agreement with the observations. Then the longitude of the ascending 
node and the longitude of the periastron from the node are determined with an 
accuracy of up to $180^\circ$ and the ephemeris of the orbits in the projection 
onto the picture plane coincide. The remaining parameters remain the same, but 
the orientation of the orbital plane in the Galaxy changes. Thus, for the stars 
presented in this paper, we obtain 1, 2, 4 solutions or a family of orbits. 

For the initial estimation of the sum of the masses of the components, we used 
the "mass-spectrum-luminosity" ratio according to the handbook ~\citep{Allen} 
and data from the WDS catalog. However, there are no spectral types for many 
stars in the WDS catalog. Therefore one had to use different sources, which do not 
always give an unambiguous result. At present, the best way to get mass estimates 
is to use evolutionary track diagrams that link color index and magnitude 
~\citep{2000A&AS..141..371G, 2012MNRAS.427..127B}. 

Since most of the components of visual binaries in our program turned out to be 
spectroscopic binaries, data on them are contained in the Tokovinin {\it Catalog 
of Multiple Stars} --- MSC, ~\citep{2018ApJS..235....6T}. Nevertheless, in some 
cases, our studies of dynamics lead to an excess of masses in stellar systems, 
which should be taken into account in further studies. 

The more accurate the input data, the more reliable the orbit. We determine the 
errors of each orbital parameter by the total influence of the errors of all 
initial parameters. Since this effect is asymmetric with respect to the obtained 
solution, we present two errors that correspond to the maximum and minimum values 
of the given orbital element. The sum of the masses of the components is both the 
initial and the refined parameter, it is functionally related to parallax, so we 
fix it. 

The main advantage of this method is that it is possible to control the quality of 
the solution obtained, because to determine the apparent movement parameters, we 
do not use all available observations, but only the basis. You can also check the 
consistency of the source data with each other. This fundamentally distinguishes it 
from methods where the main criterion is agreement with positional observations, 
which always make it possible to obtain an orbit, but work poorly on a short arc. 

Naturally, the AMP-method has limitations, and its application requires an 
individual approach. First of all, the impossibility of reconciling all the 
initial data is due to the fact that the internal subsystems can distort the 
AMPs, as well as the accepted stellar masses. In this case, the method must not 
be applied.

In those cases when it is possible to obtain an orbit by the AMP- method, it is 
currently more correct than other orbits, since it is based on a whole complex 
of observation results of both an astrometric and astrophysical nature. 

\section{Results}

The purpose of this work is to summarize many years of work, to systematize the 
results obtained earlier and to add new orbits. 

We revised all orbits obtained over 40 years, compared its with modern observations 
from the WDS catalog, with CCD observations on a 26-inch refractor in Pulkovo for 
2003-2019  ~\citep{Izm2010, 2016AstBu..71..225I, 2020AN....341..762I} and with the 
relative positions and motions calculated by us from high-precision data from the 
Gaia DR2 ~\citep{2018A&A...616A...1G}. 

Most of the objects of our study have separations from $3''$ to $39''$, except for 
a few closer binaries that are not included in the Pulkovo Observatory observational 
program, and wider pairs. Basically, these are dwarfs of spectral types F, G and K 
from the nearest vicinity of the Sun. A total of 67 pairs included in 64 visual 
binary and multiple systems were considered. Among them are three visual triples 
(ADS 48 ABF, ADS 7034 ABC, and ADS 10288 ABC), for which we have determined both 
inner and outer orbits. An unambiguous AMP-solution was obtained for 33 pairs, 2 
solutions for 14 pairs, 4 solutions for 3 pairs, families of AMP-orbits were 
calculated for 17 pairs. 

For 29 stars, the orbits have been improved. Most often, the improvement consisted 
in using a more accurate parallax from the Gaia DR2 catalog and a homogeneous series 
of Pulkovo CCD observations instead of heterogeneous ones. In this paper, we also 
publish the orbits ADS 895 and HIP 12706 obtained for the first time. Orbits of 16 
stars were completely recalculated from the Gaia DR2 data (positions, proper motions, 
parallaxes and radial velocities), and the first orbits of seven stars were obtained. 

In 20 out of 67 cases, the quasi-instantaneous motion according to the Gaia DR2 data 
contradicts the average motion according to all-world observations. This is grounds 
for suspecting the presence of an additional satellite. Therefore to improve our 
orbits we used only Gaia DR2 parallaxes, new data from the literature, and elongated 
Pulkovo series. See, for example, ~\citep{2020Ap....tmp...69S, 2021ARep...65..209R, 
2021RAA....21..291K}. 

On the other hand, 24 pairs are known to have confirmed inner subsystems. Usually, 
short-period spectral satellites did not affect the AMP-values, and only increased 
their errors. The satellite's motion was taken into account when calculating the 
outer orbit, where it possible.

The results are presented \href{http://izmccd.puldb.ru/vds.htm}{on the website izmccd.puldb} 
in the form of spreadsheets and an Appendix, which includes comments on each star and graphs 
illustrating the comparison of ephemeris with observations, as well as the region of 
stable solutions for families. 

Here we give only Table 1, which presents data characterizing the components of the 
stellar pairs studied, series of observations and obtained solutions for each star. 

\begin{table}
\caption{Identifiers, WDS data on the studied stars, characteristics of the series of observations and the obtained results}
\label{tabl}
\begin{center} 
\begin{adjustbox}{width=1.0\textwidth}
\begin{tabular}{|c|c|rl|c|rr|ll|ll|cc|r|cc|rr|c|c|c|}
\hline N & HIP & ADS & Comp. & WDS & $m_1$ & $m_2$ & $Sp1_W$ & $Sp2_W$ & $Sp1_T$ & $Sp2_T$ & $T1_W$ & $T2_W$ & $n_W$ & $T1_P$ & $T2_P$ & $n_P$ & $n_C$ & $k_0$ & NOTE & Ref. \\
  1 &     2         &     3   &   4   &      5      &   6   &    7    &   8    &   9      &   10   &   11   &   12  &  13  &   14   &   15  &   16  &   17 &  18  &  19  &  20      &   21   \\
  \hline
  1 &     50        &     -   &       &  00006-5306 &  6.55 &   9.85  &  G0IV  &   -      &   -    &   -    &  1836 & 2015 &    27  &    -  &   -   &   -  &   -  &   4  &  C       & [1] \\
  2 &    473        &     48  &  AB   &  00057+4549 &  8.98 &   9.15  &  K6    &  M0      &  K6V   &  K6V   &  1876 & 2015 &   399  &  1961 &  2019 &  148 &  056 &   2  &  CVG     & [2,3] \\
  3 &    473/428    &     48  &  AB-F &  00057+4549 &  8.98 &  10.19  &  K6V   &  M2e     &  K6V   &  K8V   &  1897 & 2007 &    37  &  1968 &  1995 &  117 &  000 &   F  &  VWM     & [2] \\
  4 &   1475        &    246  &  AB   &  00184+4401 &  8.13 &  11.04  &  M1V   &  M3.5V   &  K8V   &  M0V   &  1860 & 2015 &   123  &  1994 &  2019 &  007 &  160 &   1  &  V(C)    & [4] \\
  5 &   2844        &    497  &  AB   &  00360+2959 &  8.39 &   9.05  &  G2V   &  G7V     &  G7V   &  G6V   &  1824 & 2015 &   134  &  1971 &  2019 &  087 &  066 &   1  &  MVWÑ    &   [0] \\
  6 &   5110        &    895  &       &  01055+1523 &  9.24 &   9.93  &  K0    &          &  K2.5V &  K7V   &  1829 & 2007 &    98  &  1960 &  2019 &  046 &  093 &   2  &  V(C)WM  &   [0] \\
  7 &  12706        &     -   &  AB   &  02433+0314 &  3.54 &   6.18  &  A1V   &          &   -    &   -    &  1825 & 2014 &   235  &   -   &   -   &   -  &   -  &   2  &  CO(M)W  &   [0] \\
  8 &  12777        &   2081  &  AB   &  02442+4914 &  3.89 &   9.14  &  F7V   &  M1.5    &  F6V   &  K8V   &  1782 & 2013 &    76  &   -   &   -   &   -  &   -  &   1  &  VWG     &   [0] \\
  9 &  12780/779    &   2098  &  AB   &  02442-2530 &  7.5  &   8.1   &  G3V   &          &  G2V   &  K0V   &  1835 & 2013 &    29  &   -   &   -   &   -  &   -  &   F  &  VWM     &  [1] \\
 10 &  15058        &   2416  &       &  03140+0044 &  8.14 &   8.17  &  F8    &          &  G9V   &  K0V   &  1831 & 2012 &   167  &   -   &   -   &   -  &   -  &   2  &  C(M)    &   [1] \\
 11 &  15220        &   2427  &  AB   &  03162+5810 & 10.30 &  11.38  &  M2V   &          &  M0V   &  M0V   &  1914 & 2013 &    82  &  1971 &  2019 &  100 &  149 &   1  &  V(C)G   &   [0]  \\
 12 &  17129        &   2668  &  A-Bb &  03401+3407 &  7.52 &   7.60  &  F9V   &          &  F9V   &  F4V   &  1823 & 2012 &   230  &  2003 &  2019 &  000 &  045 &   2  &  (C)MW   &   [5] \\
 13 &  17666        &   2757  &  AB   &  03470+4126 &  8.20 &   8.82  &  K1V   &  K2V     &  G8V   &  K1V   &  1822 & 2011 &   129  &  1960 &  2019 &  099 &  058 &   1  &  VWM     &  [4] \\
 14 &  20390        &   TTau  &  NS   &  04220+1932 &  5.5  &   8.1   &  G5V   &          &   -    &   -    &  1990 & 2002 &    69  &   -   &   -   &   -  &   -  &   4  &  CWM     &   [0] \\
 15 &  32609        &   5436  &  AB   &  06482+5542 &  6.28 &   6.34  &  dF5   &  dF6     &  F6V   &  F5V   &  1821 & 2014 &   126  &  1962 &  2019 &  033 &  082 &   2  &  MVW     &   [0] \\
 16 &  33287        &   5570  &       &  06555+3010 &  8.72 &   8.97  &  G0    &          &  K0V   &  K0V   &  1831 & 2015 &   110  &   -   &   -   &   -  &   -  &   2  &  C(M)WO  &   [1] \\
 17 &  35550        &   5983  &       &  07201+2159 &  3.55 &   8.18  &  A9III &  K3V     & (F6V)  &  K1V   &  1822 & 2013 &   254  &  1972 &  2019 &  108 &  074 &   1  &  MCVW    &   [6]  \\
 18 &  40527/32     &   6646  &  AB   &  08165+7930 &  8.40 &   8.64  &  G0    &          &  F3V   &  F8V   &  1832 & 2005 &    68  &  1962 &  2003 &  041 &  000 &   F  &  MVW     &   [0]          \\
 19 &  41184/81     &   6783  &       &  08243+4457 &  7.79 &   9.39  &  G0    &          &  G2V   &  K1V   &  1830 & 2012 &    41  &  1996 &  2006 &  017 &  000 &   F  &  VWG     &   [0]          \\
 20 &  43426        &   7034  &  AB   &  08508+3504 &  7.41 &   7.48  &  F8    &          &  G0V   &  F9V   &  1821 & 2012 &   125  &  1962 &  2019 &  020 &  066 &   1  &  VWG     &   [7] \\
 21 &  43426        &   7034  &  AB-C &  08508+3504 &  7.41 &  11.69  &  F8    &          &  G0V   &  K6V   &  1941 & 2005 &     4  &   -   &   -   &   -  &   -  &   F  &  VWMG    &   [7] \\
 22 &  45343/* &   7251  &  AB   &  09144+5241 &  7.79 &   7.88  &  M0V   &  M0V     &  K7V   &  K7V   &  1821 & 2015 &   449  &  1963 &  2019 &  185 &  075 &   1  &  CVW     &    [8] \\
 23 &  48429        &   7551  &       &  09524+2659 &  9.12 &   9.50  &  K0    &          &  G7V   &  G9V   &  1830 & 2014 &   134  &  1972 &  2019 &  009 &  120 &   1  &  (C)WG   &   [0]          \\      
 24 &  48804        &   7588  &       &  09572+4554 &  8.89 &   9.75  &  G0    &          &  G0V   &  G6V   &  1828 & 2013 &    46  &  1971 &  2019 &  011 &  107 &   1  &  VWG     &   [7]  \\
 25 &  50583        &   7724  &  AB   &  10200+1950 &  2.37 &   3.64  &  K0III &          &   -    &   -    &  1782 & 2015 &   834  &  1992 &  2008 &  020 &  020 &   2  &  VW(C)   &   [9]  \\
 26 &    -          &   8002  &       &  10596+2527 &  8.57 &   9.22  &  K2    &  K5      &  K3V   &  K4V   &  1899 & 2014 &   102  &  1970 &  2018 &  061 &  056 &   1  &  CV      &   [10]  \\
 27 &  54407        &   8065  &       &  11080+5249 &  7.65 &   9.03  &  F8V   &          &  F8V   &  G7V   &  1830 & 2010 &    74  &  1970 &  2018 &  024 &  053 &   1  &  VWG(C)  &   [0]          \\
 28 &  54952        &   8100  &  AC   &  11152+7329 &  7.77 &  11.34  &  K5    &          &  K3V   &  M0.5V &  1858 & 2011 &    66  &  1969 &  1999 &  041 &  000 &   2  &  (C)W    &   [11]  \\
 29 &  56622        &   8236  &       &  11366+5608 &  7.73 &   8.17  &  G5    &  K7V     &  G2V   &  G7V   &  1828 & 2007 &   116  &  1962 &  2019 &  061 &  079 &   1  &  MVW(C)  &   [5]  \\
 30 &  56809        &   8250  &  AB   &  11387+4507 &  6.53 &   8.23  &  G0V   &          &  G0V   &  K2V   &  1782 & 2015 &   137  &  1969 &  2019 &  048 &  107 &   1  &  MVWC    &   [10] \\
 31 &  60831/32     &   8561  &       &  12281+4448 &  7.49 &   8.08  &  F9V   &          &  F8V   &  G3V   &  1791 & 2012 &    99  &  1971 &  2007 &  025 &  008 &   2  &  VWG     &   [7]  \\
 32 &  62561        &   8682  &  AB   &  12492+8325 &  5.29 &   5.74  &  A1IIIsh  &       &   -    &   -    &  1820 & 2011 &    68  &  1969 &  2007 &  023 &  000 &   F  &  MVW     &   [0]   \\
 33 &  64405        &   8814  &       &  13120+3205 &  7.40 &   7.64  &  F6V   &          &  F4V   &  F3V   &  1843 & 2014 &   255  &  1985 &  2004 &  032 &  000 &   2  &  VW(C)G  &   [0]   \\      
 34 &  65011/12     &   8861  &  AB   &  13196+3507 &  9.62 &  11.90  &  M0.5V &    M3V   &  K7V   &  K9V   &  1827 & 2012 &    53  &  1971 &  2019 &  081 &  215 &   2  &  (C)MW   &   [12]  \\
 35 &  66195        &   8959  &       &  13341+6746 &  9.26 &   9.56  &  G1V   &          &  G3V   &  G5V   &  1832 & 2007 &    47  &  1971 &  2019 &  051 &  137 &   1  &  VW(C)G  &   [0]          \\
 36 &  67422        &   9031  &       &  13491+2659 &  7.36 &   8.15  &  K4V   &    K6V   &  K3V   &  K5V   &  1823 & 2015 &   875  &  1962 &  2019 &  097 &  150 &   1  &  CG      &   [0]          \\
 37 &  67871        &   9048  &       &  13540+3249 &  8.63 &   8.97  &  F8    &          &  F9V   &  G0V   &  1823 & 2015 &    66  &  1962 &  2005 &  034 &  000 &   1  &  VWG     &   [7]  \\
 38 &  68588        &   9090  &       &  14024+4620 & 10.05 &  10.26  &  M2    &    M2.5  &  K8V   &  K7V   &  1889 & 2015 &   112  &  1962 &  2018 &  077 &  066 &   1  &  V(C)G   &   [0]          \\
 39 &   --          &     -   &       &  14051+4913 & 11.80 &  11.98  &  K4/5  &          &  G7V   &  G7V   &  1902 & 2016 &    18  &  1969 &  1975 &  022 &  000 &   F  &  VWG     &   [7]  \\
 40 &  69442        &   9167  &       &  14131+5520 &  9.06 &   9.42  &  K2    &          &  K1V   &  K1.5V &  1831 & 2013 &   202  &  1971 &  2019 &  073 &  073 &   1  &  M(C)VW  &   [0]          \\      
 41 &  69751        &   9192  &       &  14165+2007 &  6.47 &   8.42  &  F6V   &          &  F5V   &  G7V   &  1830 & 2015 &   157  &  2003 &  2019 &  000 &  119 &   1  &  CVW(M)  &   [1]  \\
 42 &  71782        &   9346  &  AB   &  14410+5757 &  7.53 &   8.32  &  K0IV  &    G5IV  &  G9V   &  G4V   &  1830 & 2015 &    97  &  1979 &  2019 &  036 &  078 &   1  &  (M)VW   &   [18]  \\
 43 &  71876        &   9357  &       &  14421+6116 &  6.33 &   9.16  &  F4V   &          &  F4V   &  K1V   &  1832 & 2015 &    46  &  2004 &  2019 &  000 &  040 &   1  &  VWG(C)  &   [7]  \\
 44 &  73846        &   9497  &  AB   &  15055-0701 &  8.09 &   8.76  &  G0    &          &  F6V   &  G1V   &  1873 & 2014 &    83  &   -   &   -   &   -  &   -  &   1  &  CW(M)   &   [1]  \\
 45 &  74666/74     &   9559  &  AB   &  15155+3319 &  3.56 &   7.89  &  G8III &      & (K3V)  &  G1V   &  1780 & 2015 &   109  &  1994 &  2005 &  010 &  000 &   F  &  VWG     &   [0]      \\
 46 &  75809/29     &   9696  &  AB   &  15292+8027 &  6.64 &   7.30  &  G0IV-V  &        & (G3V)  &  G8V   &  1823 & 2010 &    87  &  1969 &  2004 &  044 &  000 &   F  &  VWG     &   [0]          \\
 47 &  80349        &  10044  &       &  16242+3702 &  8.43 &   8.79  &  K0      &        &  K1V   &  G3V   &  1823 & 2009 &    99  &  1962 &  2019 &  026 &  255 &   F  &  MVW     &   [0]          \\
 48 &  83020        &  10288  &  AB   &  16579+4722 &  7.93 &  10.85  &  K0      &        &  K2V   &  M0V   &  1908 & 2007 &    33  &  1993 &  2019 &  010 &  150 &   1  &  V(C)G   &   [0]          \\
 49 &  83020/06     &  10288  &  AB-C &  16579+4722 &  7.93 &   8.05  &  K0V     &        &  K2V   &  K1.5V &  1823 & 2011 &    38  &  1993 &  2019 &  011 &  000 &   F  &  MVWG    &   [0]          \\
 50 &  83451/54     &  10329  &       &  17033+5935 &  8.76 &  10.34  &  K4V     &        &  K2V   &  K7V   &  1830 & 2006 &    54  &  1970 &  2019 &  022 &  050 &   1  &  VWG     &   [0]          \\
 51 &  83608        &  10345  &  AB   &  17053+5428 &  5.66 &   5.69  &  F7V     &        &  F5V   &  F5V   &  1779 & 2015 &   794  &  1965 &  2019 &  019 &  150 &   1  &  CW      &   [12] \\
 52 &  83988/96     &  10386  &  AB   &  17102+5430 &  8.85 &   9.21  &  K6V     &  K6V   &  K4V   &  K6V   &  1830 & 2012 &    42  &  1961 &  2019 &  027 &  033 &   1  &  VW      &   [4] \\
 53 &  86614/20     &  10759  &  AB   &  17419+7209 &  4.60 &   5.59  &  F5IV    &  F8V   &  F5V   &  F7V   &  1800 & 2015 &   172  &  1980 &  2005 &  076 &  003 &   F  &  MVW     &   [0]   \\
 54 &  88136/27     &  11061  &  AB   &  18002+8000 &  5.70 &   6.00  &  F7V     &  F7V   &  F7V   &  F6V   &  1782 & 2014 &   146  &  1964 &  2006 &  080 &  002 &   F  &  MVW     &   [0]    \\
 55 &  91768/72     &  11632  &  AB   &  18428+5938 &  9.11 &   9.96  &  M4      &  M4.5  &  K9V   &  K9V   &  1831 & 2015 &   608  &  1961 &  2019 &  189 &  158 &   1  &  CV      &   [13] \\
 56 &  93873/99     &  GL745  &       &  19072+2053 & 10.95 &  10.99  &  M2V     &  M2V   &  K9V   &  K9V   &  1897 & 2012 &    17  &  1996 &  2007 &  009 &  000 &   F  &  VWG     &   [0]          \\
 57 &  94336        &  12169  &  AB   &  19121+4951 &  6.54 &   6.67  &  G3V     &  G3V   &  G2V   &  G3V   &  1819 & 2015 &   286  &  1961 &  2019 &  096 &  157 &   1  &  VW      &   [4]  \\
 58 &  96895/901    &  12815  &  AB   &  19418+5032 &  6.00 &   6.23  &  G1.5V   &        &  G2V   &  G2V   &  1800 & 2014 &   582  &  1960 &  2007 &  162 &  000 &   F  &  VWGM    &   [0]          \\
 59 &  97222        &  12889  &  AB   &  19456+3336 &  8.47 &   8.58  &  K3V     &        &  K1.5V &  K1.5V &  1828 & 2015 &   450  &  1995 &  2019 &  008 &  103 &   1  &  CV      &   [14]  \\
 60 &  97295        &  12913  &  AB   &  19464+3344 &  5.06 &   9.25  &  F5V     &        &  F6V   &  K2V   &  1822 & 2014 &   138  &  1995 &  2019 &  009 &  062 &   2  &  VW      &   [14]  \\
 61 &  97292        &  GL767  &  AB   &  19464+3201 & 10.38 &  11.15  &  M0.5V   &  M2V   &  K8V   &  M0V   &  1935 & 2015 &    81  &  1971 &  2019 &  035 &  311 &   1  &  (M)V(C) &    [15]  \\
 62 & 104214/17     &  14636  &  AB   &  21069+3845 &  5.20 &   6.05  &  K5V     &  K7V   &  K5V   &  K6V   &  1753 & 2015 &  1672  &  1958 &  2019 &  328 &  254 &   1  &  CV      &   [16]  \\
 63 &    -          &  14878  &  AB   &  21200+5259 &  7.71 &   7.87  &  F8V     &        &  G0V   &  G8V   &  1828 & 2015 &   100  &  1985 &  2005 &  025 &  000 &   F  &  VW      &   [10]  \\
 64 & 105502        &  14909  &  AB   &  21221+1948 &  4.20 &   9.3   &  K0.5III &        &   -    &  K0V   &  1780 & 2013 &    75  &  1994 &  2005 &  009 &  000 &   F  &  MVW     &   [0]          \\
 65 & 108456/61     &  15571  &  AB   &  21582+8252 &  7.00 &   7.47  &  F6IV-V  &        & (F8V)  &  G6V   &  1825 & 2014 &   157  &  1960 &  2003 &  102 &  000 &   2  &  CMVW    &   [17] \\
 66 & 108917        &  15600  &  Aa-B &  22038+6438 &  4.45 &   6.40  &  A3m     &        & (F3V)  &  F7V   &  1779 & 2015 &   276  &  1963 &  2019 &  089 &  149 &   4  &  MW      &   [0]          \\
 67 & 118281        &  17149  &  AB   &  23595+3343 &  6.46 &   6.72  &  F8V     &        &  F8V   &  F8V   &  1777 & 2015 &   606  &  1980 &  2017 &  011 &  047 &   1  &  V(C)WG  &   [0]          \\
 \hline
\end{tabular}
\end{adjustbox}
\end{center}

 {\scriptsize Note: * --- ADS 7251 B = HIP 120005. Designations are given in 
the text. The last column contains a link to the article in which this orbit 
was obtained ([0] --- if in this article; [1]=~\citep{2017AstL...43..316K}; 
[2]=~\citep{2020AstBu..75..425K}; [3]=~\citep{2001AstL...27..391K}; [4]= 
~\citep{2021ARep...65..209R}; [5]=~\citep{2018AstL...44..787K}; [6]=
~\citep{2007IAUS..240..119S}; [7]=~\citep{2020AstL...46..555K}; [8]=
~\citep{2020Ap....tmp...69S}; [9]=~\citep{2014ARep...58...30R}; [10]=
~\citep{gao5st}; [11]=~\citep{gr2}; [12]=~\citep{gao9st}; [13]=~\citep{gao42st}; 
[14]=~\citep{2017ARep...61..206R}; [15]=~\citep{2015AstL...41..417K}; 
[16]=~\citep{2017Ap.....60..507S}; [17]=~\citep{2006Ap.....49..397G}; 
[18] = ~\citep{2021RAA....21..291K}).}

\end{table} 

Designations: Columns 1-3 give numbers of pairs (1 --- in order, 2 --- according 
to the Hipparcos catalog ~\citep{HIP1997}, 3 --- according to the ADS catalog 
~\citep{Aitken1932}), column 4 gives the components designations, 5 is the WDS 
number ~\citep{WDS}, 6--9 are the magnitudes and spectral types of the components 
according to WDS, 10 and 11 are the spectral types of the components estimated by 
us according to the effective temperature Teff from the Gaia DR2 data ~\citep{2018A&A...616A...1G} 
and monograph ~\citep{1981MoIzN....S....A} under the assumption that this component 
is a dwarf (otherwise, the result is given in parentheses); 12 and 13 are the beginning 
and end of all WDS observations (version of 2016), 14 is the total number of 
observations in the WDS, 15 and 16 are the beginning and end of Pulkovo observations, 
17 is the number of Pulkovo photographic observations, 18 is the number of Pulkovo 
CCD observations, 19 is the number of AMP-orbits (k0) or F --- family, 20 is code 
of one letter each: M is the presence of an inner subsystem (in brackets, if the 
satellite is assumed by us, but not yet confirmed), V is the presence of the radial 
velocity from observations, C is the presence of the radius of curvature from the 
observations of the basis (in parentheses, if over the entire arc, but not determined 
from the short basis), W is a wide pair ($a > 100$ au), O is AMP-orbit improved by 
the ORBITX program ~\citep{1992ASPC...32..573T}, G is AMPs of the final chosen orbit 
calculated from observations and proper motions of Gaia DR2; column 21 is a reference 
to the article in which this orbit was obtained. 

Note that photographic observations with the 26-inch refractor at Pulkovo ended 
in 2007 and CCD observations began in 1996, but in this work we have used only 
systematic CCD observations since 2003 to 2019. 

Table 2 presents the initial data for obtaining the AMP-orbit. Namely: the 
apparent motion parameters and additional parameters, as well as their errors.

Table 3 lists the orbital elements for 50 pairs, their errors depending on the initial 
parameters, as well as the orientation of the obtained orbits in the galactic coordinate 
system and the average-weight values (О-С), corresponding to the entire series and to 
the Gaia DR2 observation. The assignment of weights to observations is explained in the 
article ~\citep{2017AstL...43..316K}. The initial data errors correspond to $1\sigma$, 
then the orbital elements fall into the specified error range with a probability of 
$68\%$. 

Table 4 shows the orbital elements for 17 pairs (families) corresponding to singular 
points, i.e. having a minimum period ($\beta=0^\circ$), a minimum eccentricity, and 
limiting periods for reliable orbits of the families. 

Here we should note the peculiarity inherent in the algorithm for determining orbits 
by position and velocity. See the monographs ~\citep{1968itta.book.....S} and 
~\citep{HT2007}. 

First of all, we obtain the semi-major axis of the orbit $a$ according to the energy 
integral by the formula: 

\begin{equation}
\label{four} \frac{1}{a}=\frac{2}{r}-\frac{v^2}{k^2}
\end{equation}
For values of $\beta$ close to $\beta_{max}$, when the orbit is close to parabolic, 
the error in determining $a$ is large. Then calculating the ephemeris from the found 
orbit, we get an erroneous value of $r$: 

\begin{equation}
\label{five} r=a(1-e\cos{E})
\end{equation}

Here $e$ is the eccentricity, $E$ is the eccentric anomaly. In addition to the 
inaccurate value of $a$, the value of $r$ is affected by the position that the 
satellite occupies in orbit at the instant $T_0$. It is obvious that a greater 
uncertainty is obtained near the periastron. In Table 4, we include only reliable 
solutions when the cycle of calculating the orbit and its ephemeris is closed 
and all the orbits of the family fit the observed series well. 

In the Appendix (\href{http://izmccd.puldb.ru/vds.htm}{on the site izmccd.puldb}) 
comments and graphics are given for each star. The comments briefly describe the 
history of the study of the system and the rationale for this result. The graphs 
show series of observations, orbit ephemeris, and the direction of motion 
according to Gaia DR2 data at the instant of 2015.5. 

The following dependences are presented: $\rho(t)$, $\theta(t)$ and $y(x)$. A graph 
in the picture plane $y(x)$ is sometimes presented in two forms: a fragment of an 
arc covered by observations, and a complete orbit during the entire period. Then 
you can see how small the observed arc is. 

For all families, the functional dependence $lg(a)=f(e)$ is additionally presented 
as a graph and the region of possible influence of the Galaxy is determined. For 
visual triple stars (ADS 48, ADS 7034 and ADS 10288), the stability region is also 
determined as it was done in the work  ~\citep{2020AstL...46..555K}. 

\section{Conclusion}

The orbits of 67 wide pairs of visual binary and multiple stars included in 64 
systems are presented. Orbits were unambiguously obtained for 33 pairs, among 
them 22 pairs have a semi-major axis of the orbit of more than 100 au. and periods 
from 600 to 6000 years. For such wide pairs, the AMP- method, in contrast to other 
methods, makes it possible to obtain more reliable orbits. 

For 40 years, each researcher from the group of A.A.Kiselev has being developed 
his own algorithms for determining the AMP-orbits, corresponding to the features 
of the objects of interest to him, depending on the available initial data. To 
achieve research uniformity and ease of use, we have created a single program 
that, using the data in Table 2 for each pair, calculates orbit elements with 
errors, orientation, ephemeris, and deviations to observations (Tables 3 and 4, 
as well as data for graphs). This program can serve as a tool for determining 
the AMP-orbits of yet unexplored visual binaries of the Pulkovo program as the 
missing data become available. 

\section*{Acknowledgments}

This article uses catalogs data from the WDS ~\citep{WDS} and Gaia DR2 
~\citep{2018A&A...616A...1G}, the authors are grateful to their creators. 
This work was supported by the Russian Foundation for Basic Research 
(grant No. 20-02-00563 A).

\bibliography{references}

\end{document}